\documentclass{iau}

\usepackage{amsmath}
\usepackage{graphicx}
\usepackage{multirow}

\def \jnl{}

\def\aap{\jnl{A\&A}}                
\def\apj{\jnl{ApJ}}                 
\def\aj{\jnl{AJ}}                 
\def\apjs{\jnl{ApJS}}                 
\def\araa{\jnl{ARAA}}                 
\def\mnras{\jnl{MNRAS}}             
\def\nat{\jnl{Nature}}              

\begin{document}

\lefttitle{Efrain Gatuzz}
\righttitle{Bridging observations and simulations}

\jnlPage{1}{7}
\jnlDoiYr{2025}
\volno{397}
\doival{10.1017/xxxxx}

\aopheadtitle{Proceedings IAU Symposium}
\editors{C. Sterken,  J. Hearnshaw \&  D. Valls-Gabaud, eds.}

\title{Bridging observations and simulations: a machine learning approach to galaxy clusters}

\author{Efrain Gatuzz}
\affiliation{Max-Planck-Institut f\"ur extraterrestrische Physik (MPE), Gie{\ss}enbachstra{\ss}e 1, 85748 Garching, Germany\\
  \email{egatuzz@mpe.mpg.de}}

\begin{abstract}
The intracluster medium (ICM) records the history of galaxy clusters through its complex dynamical properties. 
To effectively interpret these properties, robust methods are needed to compare observational data with theoretical models. 
We present a novel machine learning framework for comparing ICM line-of-sight velocity maps derived from X-ray observations. 
Our approach uses convolutional and Siamese neural networks to identify similarities between different kinematic fields. 
We outline the architecture of this framework and perform a series of sanity checks to validate its performance. 
These checks demonstrate the model’s ability to correctly identify and quantify kinematic features, establishing a powerful new tool for future comparative studies of the ICM. 
\end{abstract}


\maketitle

\vspace{-18pt} 
\section{Introduction}\label{sec_int}  
The ICM, a hot, diffuse plasma, constitutes the dominant baryonic component of galaxy clusters \citep{and10,dai10,kra12}. 
Its complex dynamics are driven by various astrophysical processes, including cluster mergers, AGN feedback, and large-scale structure formation, which generate turbulent motions, bulk flows, and coherent velocity patterns such as gas sloshing \citep{asc06,vaz18,zuh18,bam19,ich19}.

Recent advances in X-ray spectroscopy, particularly through the calibration technique introduced by \citet{san20}, have enabled direct measurements of ICM line-of-sight (LOS) velocities with high precision. 
Applying this method to clusters such as Perseus, Virgo, and Ophiuchus has revealed a range of kinematic phenomena, including AGN-driven outflows and merger-induced bulk flows. 
Interpreting these observations requires comparison with predictions from cosmological hydrodynamical simulations \citep{gas14,zhu14,moh19}. However, the high dimensionality and inherent complexity of both observational and simulated datasets pose a significant challenge for direct comparison. 

In this work, we present a novel methodology that employs machine learning to address this problem. 
Specifically, we use Convolutional Neural Networks (CNNs) and Siamese architectures to construct a scalable framework for comparing observed LOS velocity maps with synthetic counterparts derived from simulations. 
This approach builds upon successful applications of similar techniques in other areas of astrophysics \citep{hue15,dom18,pea19,bru19,zha22}, and is designed to identify both kinematic and morphological similarities between observed and simulated ICM structures. 
In the following sections, we outline the core components of this method.

\section{Galaxy Cluster Velocity Maps}\label{sec_vel_map_dat}
To test our machine learning framework, we used LOS velocity maps from several galaxy clusters: Virgo \citep{gat22a,gat23b}, Centaurus \citep{gat22b,gat23d}, Ophiuchus \citep{gat23a,gat23e}, and A3266 \citep{gat24a,gat24b}. 
These maps were derived from {\it XMM-Newton} EPIC observations, processed using a standardized data reduction and analysis pipeline.
Briefly, the observations were reduced with SAS (version 19.1.0), including the selection of single-pixel events and the filtering of flaring time intervals. 
A critical aspect of the process involved the application of updated calibration files and the use of instrumental background lines to achieve a LOS velocity precision better than 100 km/s at the Fe-K line, following the method described in \citet{san20}. 
Point sources, including central AGN, were identified and excluded from the analysis.
Spectral maps were generated by fitting X-ray spectra extracted from dynamically sized elliptical regions across each observation. 
The ICM emission was modeled using an \texttt{apec} thermal plasma component \citep{fos19}, with additional components introduced in cases requiring more complex modeling, such as the multi-temperature structure observed in the Centaurus cluster.
This procedure, which accounts for Galactic absorption and instrumental background contributions, yielded the LOS velocity maps that serve as inputs for our machine learning framework. 
Further details on the reduction and spectral analysis techniques can be found in the original references.

\section{Machine Learning Methodology}
We present a deep learning framework based on a Siamese Convolutional Neural Network (CNN) designed to compare velocity maps of the ICM. 
The objective is to identify kinematically similar fields by learning a robust representation, or \textit{embedding}, for each input velocity map.
The Siamese CNN consists of two identical branches with shared weights, facilitating similarity learning. Each branch functions as a feature extractor, composed of multiple convolutional layers that capture hierarchical spatial features, followed by max-pooling layers to reduce spatial dimensions. 
After the final convolutional block, a flattening layer and a fully connected dense layer convert the feature maps into a compact, fixed-length numerical vector-the embedding-which encodes the essential characteristics of the input map.

The network is trained using the triplet loss function \citep{sch15}. 
During training, the network is presented with triplets: an anchor image (A), a positive image (P) that is similar to the anchor, and a negative image (N) that is dissimilar. 
The loss function adjusts the network’s weights via back propagation to minimize the distance between the anchor and positive embeddings while enforcing a margin between the anchor and negative embeddings.
Once training is complete and the weights are optimized, the CNN branches act as fixed feature extractors. 
The observed velocity map from \textit{XMM-Newton} is passed through one branch to generate its embedding vector, while each simulated velocity map is processed through the second branch to compute its corresponding embedding. 
A critical preprocessing step ensures all images are resampled to a uniform pixel resolution consistent with the \textit{XMM-Newton} data.

To quantify similarity, we compute both the Euclidean distance and the cosine similarity between the embeddings of the observation and each simulation. 
The simulation with the smallest Euclidean distance (or, equivalently, the highest cosine similarity) is identified as the best match.

\section{Validation Using Artificial Data}\label{sec:validation} 
To assess the robustness and reliability of our Siamese CNN in identifying kinematic similarities, we conducted a series of validation tests using controlled artificial datasets. 
These tests serve as sanity checks to confirm that the network learns meaningful and distinctive representations of ICM velocity maps.

\begin{figure*}
\centering
\includegraphics[width=0.99\linewidth]{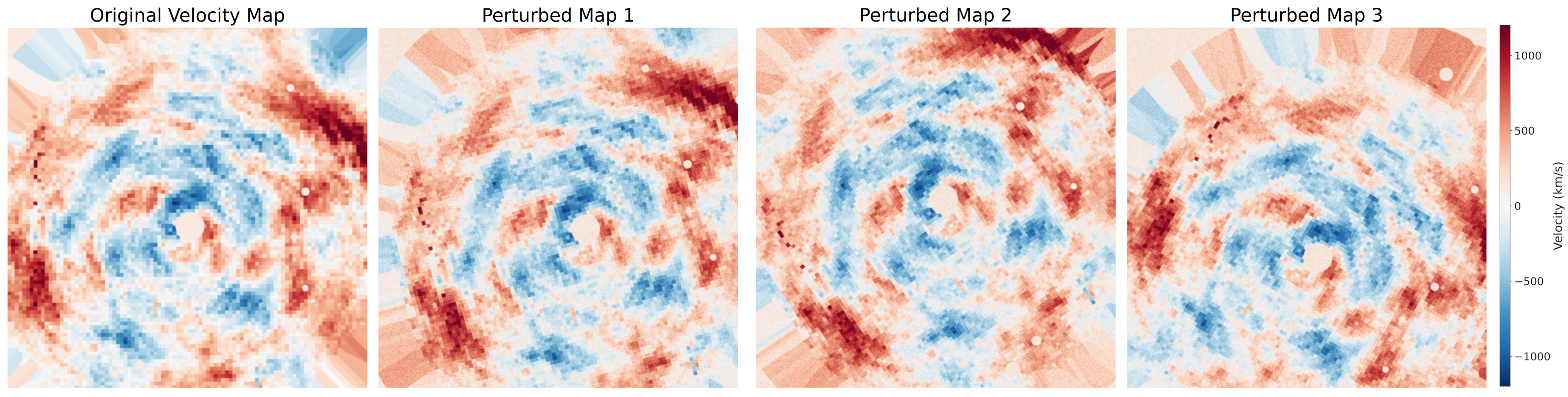}
\caption{Examples from the artificial dataset generated for the Virgo cluster. 
These perturbed velocity maps were created by applying controlled transformations (e.g., rotation, noise) to the original \textit{XMM-Newton} observation.}
\label{vir_fake_sample}
\end{figure*}

For each observed cluster-Virgo, Centaurus, Ophiuchus, and A3266-we generated a set of 1,000 artificial velocity maps. 
These maps were created by applying controlled perturbations to the original \textit{XMM-Newton} observations, including random rotations, pixel-level noise additions, and minor distortions (see Figure~\ref{vir_fake_sample}). 
We then trained the Siamese CNN on a subset of this artificial dataset.

To evaluate performance, we queried the trained model with the original (unperturbed) observational map and searched for its closest match among a validation set containing the 1000 perturbed maps along with the original. 
In every case, the model correctly identified the original observation as the best match-returning the smallest embedding distance and highest similarity score. 
This outcome confirms that the Siamese CNN has successfully learned to recognize the intrinsic kinematic patterns of the ICM, even when subjected to small but realistic variations.

\begin{figure*}
\centering
\includegraphics[width=0.65\linewidth]{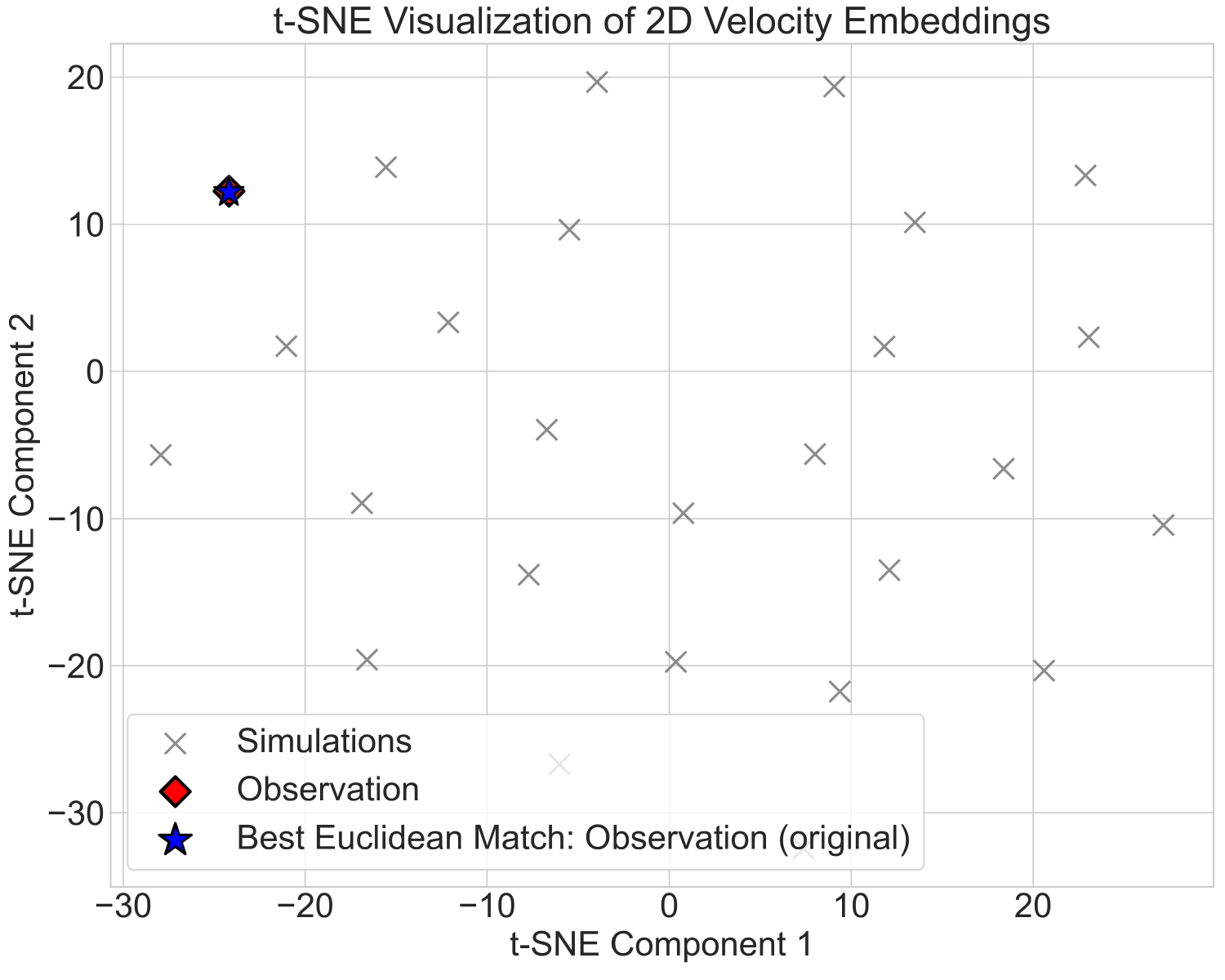}
\caption{t-SNE projection of the embedding space for the Virgo cluster. 
The plot shows that the best matching correspond to the original observation, demonstrating that the Siamese CNN learns consistent kinematic representations.}
\label{vir_tsne}
\end{figure*}

To further support these findings, we employed t-Distributed Stochastic Neighbor Embedding (t-SNE) to visualize the embedding space. 
This dimensionality reduction technique projects high-dimensional embedding vectors into a 2D space while preserving local structures. 
In these t-SNE plots, each point represents the embedding of one artificial velocity map.
As shown in Figure~\ref{vir_tsne}, the embeddings of the 1000 perturbed Virgo maps form a compact cluster tightly surrounding the embedding of the original observation. 
This visualization demonstrates that the network has learned a coherent representation of the underlying ICM kinematics-grouping similar structures together while separating dissimilar ones in the embedding space.

\section{Conclusion}
In this work, we have introduced a novel machine learning framework for the comparative analysis of ICM kinematics. 
Our approach leverages a Siamese CNN trained with a triplet loss function to learn meaningful and robust representations of velocity maps derived from X-ray observations. 
This architecture is specifically designed to identify and quantify similarities between different kinematic fields.

To validate the effectiveness of our method, we performed a series of controlled tests using artificially perturbed datasets based on real \textit{XMM-Newton} observations. 
The model consistently identified the original, unperturbed maps as the closest match, demonstrating its ability to capture the essential kinematic features. 
These results were further supported by t-SNE visualizations, which showed that embeddings of the perturbed maps clustered tightly around those of the originals-providing strong evidence that the network has learned to distinguish meaningful patterns in the data.

This framework offers a scalable and automated solution for matching observations with simulations, addressing a key challenge in the comparison of high-dimensional, complex astrophysical datasets. 
It serves as a powerful tool for linking X-ray observations with theoretical models of cluster formation and evolution.

In a forthcoming paper, we will apply this methodology to cosmological hydrodynamical simulations from the IllustrisTNG suite \citep{spr18}, providing a comprehensive analysis of how observed ICM velocity structures compare to those predicted by state-of-the-art simulations.
 
\bibliographystyle{iaulike}




\end{document}